\documentstyle[12pt]{article}
\begin{document}

\begin{titlepage}

\hskip3cm {\it  ULB--TH--98/08, FTUV/98--81, IFIC/98--82, ESI-625}

\begin{centering}

\vspace{0.5cm}

\huge{A general non renormalization theorem in the extended antifield 
formalism}\\

\vspace{.5cm}

\large{Glenn Barnich$^*$}\\

\vspace{.5cm}

Physique Th\'eorique et Math\'ematique, Universit\'e Libre de Bruxelles,
Campus Plaine C.P. 231, B-1050 Bruxelles.

\end{centering}

\vspace{2.5cm}

\begin{abstract}
In the context of algebraic renormalization, the extended antifield
formalism is used to derive the general forms of the anomaly
consistency condition and of the Callan-Symanzik equation for generic
gauge theories. A local version of the latter is used to derive
sufficient conditions for the vanishing of beta functions associated to 
terms whose integrands are invariant only up to a divergence for an
arbitrary non trivial non anomalous symmetry of the Lagrangian. 
These conditions are independent of power counting restrictions
and of the form of the gauge fixation. 
\end{abstract}

\vspace{2.5cm}

\footnotesize{$^*$Charg\'e de Recherches du Fonds National
Belge de la Recherche Scientifique.\\
New address: Universitat de Val\`encia, Departament de F\'{\i}sica 
Te\`orica, Facultad de F\'{\i}sica, C. Dr. Moliner 50, 
                E-46100 Burjassot (Val\`encia) Spain.}

\end{titlepage}

\section*{Introduction}

\def\qed{\hbox{${\vcenter{\vbox{                         
   \hrule height 0.4pt\hbox{\vrule width 0.4pt height 6pt
   \kern5pt\vrule width 0.4pt}\hrule height 0.4pt}}}$}}
\newtheorem{theorem}{Theorem}
\newtheorem{lemma}{Lemma}
\newtheorem{definition}{Definition}
\newtheorem{corollary}{Corollary}
\newcommand{\proof}[1]{{\bf Proof.} #1~$\qed$}

\def\be{\begin{eqnarray}}
\def\ee{\end{eqnarray}}
\def\G{\Gamma}
\def\D{\Delta}

Some of the perturbative proofs \cite{BlCo,KaPr,CFHP} of the 
vanishing of the $\beta$
function for Chern-Simons theory rely on two essential steps.
The first is the use of a local, non integrated version of the
Callan-Symanzik equation, where the $\beta$ functions are associated to
strictly BRST invariant terms, the second is the fact that the
Chern-Simons term is gauge, respectively
BRST invariant, only up to a total divergence. 

This article is motivated by the desire to understand this non
renormalization mechanism independently of the gauge fixation and of
power counting restrictions, so that they can be applied 
to effective field theories \cite{GoWe,Wei}. Its aim is to derive 
general conditions under which the above mechanism can be extended to
generic theories with arbitary non trivial global or local
symmetries. This is done on the one hand by using the 
extended antifield formalism with the corresponding BRST differential
to control the renormalization of the symmetries and, on the other
hand, by implementing a local version of the Callan-Symanzik equation
through an anticanonical transformation. 

The article is organized as follows. In the first section, we review
the quantum action principles, which will allow us to control the
renormalization aspects of the problem. 

The second section summarizes
the results on the extended antifield formalism and the corresponding
BRST cohomology. 

A first application in section 3 concerns the
derivation of the general form of the anomalous extended Zinn-Justin 
equation. More precisely, it is shown that by choosing suitable BRST
breaking finite counterterms at each order, the anomalous insertion at
the next order is characterized by the cohomology of the extended
BRST differential in ghost number $1$.  

In the main section, power 
counting is implemented  by a canonical transformation in the
antibracket. 
For simplicity, the case where there are no dimensionful couplig constants
and the theory is dilatation invariant is discussed next. The general
form of the integrated Callan-Symanzik equation is given. Then a local
form is derived toghether with sufficient conditions for 
the above mentioned non renormalization mechanism. As a direct application,
Chern-Simons theory is discussed. In the next part, the analysis is
generalized to the case where dilatation invariance is broken through
the presence of dimensionful coupling constants.

In the conclusion, we comment on possible extensions for this
approach. Finally, the appendix contains three technical points needed
in the main text.

\section{Quantum action principle}
In order to deal with the renormalization aspects of the problem,
we will use the quantum action principles \cite{QAP}
as pioneered in \cite{BRS} and elaborated for instance 
in \cite{PiRo,PiSo}.
Let $S$ be the classical action of the theory. If $\G$ denotes the
renormalized generating functional for one particle irreducible
vertices, one has
\be
\G=S+O(\hbar).
\ee
Similarily, if $\D\circ\G$, respectively $\D(x)\circ\G$,
denotes the renormalized insertion of an (integrated) local polynomial into 
$\G$, one has 
\be
\D\circ\G=\D+O(\hbar).
\ee 

Let $g$ be a parameter of $S$, $\phi(x)$ the source for the 
1PI vertex functions and $\rho(x)$ an external source coupling to 
a polynomial in the fields and their derivatives. 
We will use the quantum action principle in the following forms: 
\be
{\rm (non\ linear)\ field\ variations:}\nonumber\\ 
{\delta\G\over\delta \phi(x)}
{\delta\G\over\delta \rho(x)}
=\D^{\prime\prime}(x)\circ\G,\nonumber\\ \D^{\prime\prime}(x)\circ\G
={\delta S\over\delta \phi(x)}{\delta S\over\delta \rho(x)}+O(\hbar).
\ee
\be
{\rm coupling\ constants:}\ 
{\partial\G\over\partial g}=\D\circ\G,\nonumber\\ 
\D\circ\G={\partial S\over\partial g}+O(\hbar).
\ee
This is the only input from renormalization theory that will be used
in the following.

\section{Extended antifield formalism}

The Batalin-Vilkovisky formalism \cite{Zin,BaVi,HeTe,GPS} is a tool to 
control 
gauge symmetries under renormalization for generic 
gauge theories. A well defined quantum
theory requires the introduction of a non minimal sector followed by
an anticanonical transformation to a gauge fixed basis. 
The antifields are not set to zero, but kept
as external sources, since their presence in the gauge fixed theory
allows to control the renormalization of the gauge or the global
symmetry. 
Although it is always understood in intermediate
computations, we will not make the passage to the gauge fixed basis
explicitly, because the considerations below 
rely only on the BRST cohomology
of the theory. This cohomology does not depend on the non minimal
sector \cite{Hen} and is invariant under anticanonical
transformations.

The formalism can be extended so as to 
include (non linear) global symmetries (see \cite{BHW} 
and references therein), which is achieved by coupling the 
BRST cohomology classes in negative ghost numbers with constant
ghosts. There is a further extension \cite{Bar1} to include the 
BRST cohomology classes in all the ghost numbers, which allows to 
take into account in a systematic way all higher order cohomological 
constraints due to the antibracket maps \cite{Bar}. 

Let us briefly summarize the results of \cite{Bar1}
needed in the following. The extended formalism is obtained by 
first computing a basis 
for the local BRST cohomology classes, containing 
as a subset those classes that can be obtained from the solution 
$S$ of the master equation by differentiation of $S$ with respect to an 
essential coupling constant\footnote{A set of coupling constants $g^i$ 
is essential if $\lambda^i\partial S/\partial g^i=(S,\Xi)$ for some
local functional $\Xi$ implies $\lambda^i=0$. A parametric dependence
on inessential couplings related for example to gauge fixation 
is always understood in the following. More details
on the dependence of the theory on inessential couplings will be
given elsewhere.}. 
The additional classes are then coupled
with the help of new independent coupling constants to the solution of 
the master equation. This action can then be extended by terms of
higher orders in the new couplings in such a way that, if we denote by 
$\xi^A$ both the essential couplings and the new ones, the
corresponding action $S(\xi)$ satisfies the extended master equation
\be
\frac{1}{2}(S(\xi),S(\xi))+\Delta_c S(\xi)=0.\label{1}
\ee
The BRST differential associated to
the solution of the extended master equation is 
\be
\bar s=(S(\xi),\cdot)+\Delta_c,
\ee
where $\Delta_c=(-)^Af^A(\xi){\partial^L\over\partial\xi^A}$ is at
least quadratic in the new couplings and satisfies
$\Delta_c^2=0$. Since it does not depend on the fields and the
antifields, it also satisfies
$\Delta_c(A,B)=(\Delta_cA,B)+(-)^{A+1}(A,\Delta_c B)$.
The local BRST cohomology classes contain
the generators of all the generalized non trivial symmetries of the
theory in negative ghost number, the generalized observables in ghost
number zero, and the anomalies (and anomalies for anomalies) in
positive ghost number. This is the reason why the extended master
equation encodes the invariance of the original action under all
the non trivial gauge and local symmetries, their commutator
algebra as well as the antibracket algebra of all the 
local BRST cohomology classes. 

The cohomology of $\bar s$ in the space $F$ of $\xi$ dependent local 
functionals in the fields, the antifields and their derivatives, 
over the functions in the $\xi^A$, is isomorphic to
the cohomology of 
\be
s_Q=[{\partial^R\cdot\over\partial\xi^A}f^A(\xi),\cdot]
\ee
in the space of graded
derivations ${\partial^R\cdot}/{\partial
  \xi^A}\lambda^A(\xi)$, with $\lambda^A$ a function of $\xi$ 
alone. 
If ${\partial^R\cdot}/{\partial
  \xi^B}\mu^B(\xi)$ is a $s_Q$ cocycle, the corresponding $\bar s$ cocycle is
given by ${\partial^R S(\xi)}/{\partial
  \xi^B}\mu^B(\xi)$. Furthermore, the general solution to the equations 
\be
\left\{\begin{array}{c}{\partial^R S(\xi)\over\partial
  \xi^B}\mu^B(\xi)+\bar s C(\xi)=0,\\
s_Q{\partial^R\cdot\over
  \partial\xi^B}\mu^B(\xi)=0,
\end{array}\right.\label{8}
\ee
 is 
\be
\left\{\begin{array}{c}
C(\xi)=\bar s D(\xi)+ {\partial^R S(\xi)\over\partial
  \xi^B}\nu^B(\xi),\\
{\partial^R\cdot\over\partial
  \xi^B}\mu^B(\xi)=s_Q{\partial^R\cdot\over\partial
  \xi^B}\nu^B(\xi).\end{array}\right.\label{9}
\ee

A crucial property of the extended antifield formalism is that it is
stable, in the sense that every infinitesimal deformation of the
solution of the extended master equation, characterized by
$H^{0,n}(\bar s|d)$, can be extended to a complete deformation without
any obstructions. This makes it an appropriate starting point for a
quantum theory.

\section{Anomalies to all orders}

In the following, we will omit from the notation  
the dependence on the couplings $\xi^A$.
The action principle applied to (\ref{1}) gives
\be
\frac{1}{2}(\G,\G)+\D_c\G=\hbar{\cal A}\circ\G\label{2},
\ee
where $\G$ is the renormalized generating functional for 1PI vertices
associated to the solution $S$ of the extended master equation
and the local functional ${\cal A}$ is an element of 
$F$ in ghost number $1$. 
Applying
$(\G,\cdot)+\D_c$ to (\ref{2}), the l.h.s vanishes identically
because of the 
graded Jacobi identity for the antibracket and the properties of
$\D_c$, so that one gets
the consistency condition $(\G,{\cal A}\circ\G)+\D_c{\cal A}\circ\G
=0$. 
To lowest order
in $\hbar$, this gives $\bar s {\cal A}=0$, the general solution of
which can be writen as 
\be
{\cal A}={\partial^R S\over\partial \xi^A}\sigma^A_1+\bar s \Sigma_1,
\ee
with $s_Q {\partial^R\cdot /\partial \xi^A}\sigma^A_1=0$.
If
one now defines ${S}^1=S-\hbar\Sigma_1$, the
corresponding generating functional admits the expansion 
$\G^1=\G-\hbar\Sigma_1+O(\hbar^2)$. 
This implies that $\frac{1}{2}(\G^1,\G^1)+\D_c\G^1=\hbar
{\partial^R \G^1/\partial \xi^A}\sigma^A_1+O(\hbar^2)$. On the
other hand, the quantum action principle applied to
$\frac{1}{2}(S^1,S^1)+\D_c S^1=O(\hbar)$ implies
$\frac{1}{2}(\G^1,\G^1)+\D_c\G^1=\hbar\bar{\cal A}\circ\G^1$, for a local
functional $\bar{\cal A}$. Comparing the two
expressions, we deduce that 
\be
\frac{1}{2}(\G^1,\G^1)+\D_c\G^1=\hbar
{\partial^R \G^1\over\partial \xi^A}\sigma^A_1
+\hbar^2{\cal A}^\prime\circ\G^1 
\ee
for a local functional ${\cal A}^\prime$.

Applying now
$(\G^1,\cdot)+\Delta_c$, one gets as consistency condition
\be
{\partial^R\over\partial \xi^A}[{\partial^R\G^1\over\partial \xi^B}
\sigma^B_1+\hbar{\cal A}^\prime\circ\G^1]\sigma^A_1
+(\G^1,{\cal A}^\prime\circ\G^1)
+\Delta_c{\cal A}^\prime\circ\G^1=0,
\ee
giving to lowest order 
\be
{\partial^R S\over\partial \xi^A}1/2[\sigma_1,\sigma_1]^A+\bar s 
{\cal A}^\prime=0,
\ee
 where ${\partial^R\cdot /\partial\xi^A}[\sigma_1,\sigma_1]^A=[{\partial^R 
\cdot/\partial\xi^A}\sigma^A_1,{\partial^R\cdot
/\partial\xi^B}\sigma^B_1]$, 
and is 
a $s_Q$ cocycle because of the graded Jacobi identity for the graded 
commutator. According to the previous section, the general solution to this 
equation is 
\be
{\partial^R\cdot \over\partial\xi^A}[\sigma_1,\sigma_1]^A
=[{\partial^R\cdot \over\partial\xi^A}f^A,{\partial^R\cdot \over
\partial\xi^B}\sigma^B_2], \\
{\cal A}^\prime=\bar s \Sigma_2+{\partial^R
  S\over\partial\xi^B}\sigma^B_2.
\ee 
The redefinition $S^2=S^1-\hbar^2\Sigma_2$ then allows to achieve 
\be
\frac{1}{2}(\G^2,\G^2)+\D_c\G^2=
{\partial^R \G^2\over\partial \xi^A}(\hbar\sigma^A_1+\hbar^2\sigma^A_2)
+\hbar^3{\cal A}^{\prime\prime}\circ\G^2,
\ee
for a local functional ${\cal A}^{\prime\prime}$.
The reasoning can be
pushed recursively to all orders with the result 
\be
\frac{1}{2}(\G^\infty,\G^\infty)+\D_c\G^\infty=
{\partial^R \G^\infty\over\partial \xi^A}\sigma^A,
\ee
where ${\G}^\infty$ is associated to the action 
${S}^\infty=S-\Sigma_{k=1}\hbar^k\Sigma_k$ and 
$\sigma^A=\Sigma_{k=1}\hbar^k\sigma_k$ satisfies 
$s_Q{\partial^R\cdot \over\partial\xi^A}\sigma^A=0$.

This result agrees with the one deduced in \cite{HLW,Whi} 
without the $\D_c$ operator and the associated $s_Q$ cohomology
and the one in \cite{Bar1} in the context of
dimensional regularisation. It answers the questions raised in
\cite{DPT} on higher order consistency conditions on the anomalies and
the quantum BRST cohomology (see also \cite{PT}).

In the case where $H^1(\bar s)\simeq H^1(s_Q)=0$, we see that we
can achieve 
(\ref{2}) without any anomalous insertion:
\be
\frac{1}{2}(\G^\infty,\G^\infty)+\D_c\G^\infty=0.\label{3}
\ee
If this equation holds, we say that the non trivial symmetries
encoded in the differential $\bar s$ are non anomalous.

\section{Integrated and local Callan-Symanzik equations}

\subsection{Power counting}

In the antifield formalism, power counting can be implemented canonically
through the operator 
\be
S_\eta=\int d^nx\ L_\eta=\int d^nx\ \phi^*_a(d^{(a)}+x^\mu\partial_\mu)
\phi^a,
\ee
where $\phi^a$ is a collective notation for the original fields and
the local ghosts associated to the gauge symmetries, while $d^{(a)}$ is
the canonical dimension of $\phi^a$ in units of inverse length. The  
bracket around the index $a$ means that there is no additional summation. 
We have
$(\phi^a(x),S_\eta)=(d^{(a)}+x^\mu\partial_\mu)\phi^a(x)$ 
and
$(\phi^*_a(x),S_\eta)=(n-d^{(a)}+x^\mu\partial_\mu)\phi^*_a(x)$,
so that the canonical dimension of the antifields is choosen to be
$n-d^{(a)}$. It is then straightforward to verify that for any
monomial $M(x)$ in the fields, the antifields and their derivatives of
homogeneous dimension $d^M$,
\be
(M(x),S_\eta)=(d^M+x^\mu\partial_\mu)M(x).
\ee

\subsection{No dimensionful coupling constants}

For simplicity, we assume in a first stage that the 
coupling constants $\xi^A$ as well as the inessential coupling
constants all have dimension $0$.
In this section, we also assume that there
are no anomalies, eq (\ref{3}).

\subsubsection{Integrated form of Callan-Symanzik equation}
  
Because there are non dimensionful parameters, 
all the terms of the Lagrangian 
$L$ of the solution of the extended master equation
have dimension $n$. Hence,
\be
({L},S_\eta)=(n+x^\mu\partial_\mu){L}
=\partial_\mu(x^\mu{L}).\label{4}
\ee
Upon integration, we get 
\be
(S,S_\eta)=0.
\label{bl}
\ee
Furthermore, 
$\Delta_c S_\eta=0$ because 
$S_\eta$ does not depend on $\xi^A$.
We have $(S^\infty,S_\eta)=O(\hbar)$, so that the 
quantum action principle gives
\be
(\G^\infty,S_\eta)=\hbar{\cal B}\circ\G^\infty,\label{zin}
\ee
with ${\cal B}$ an element of $F$ of ghost number $0$. 

Applying $(\G^\infty,\cdot)+\D_c$, using the graded Jacobi identity and
(\ref{3}), we get the consistency condition 
$(\G^\infty,{\cal B}\circ\G^\infty)+\D_c{\cal B}\circ\G^\infty=0$, 
which implies, to lowest
order in $\hbar$, $\bar s {\cal B}=0$ and hence 
\be
{\cal B}
=\frac{\partial^R S}{\partial \xi^A}\beta^A_1+\bar s\Xi_1.
\ee
 
According to the quantum action principle, we can replace 
$\frac{\partial^R S}{\partial \xi^A}\beta^A_1\circ\G^\infty$ by 
$\frac{\partial^R \G^\infty}{\partial \xi^A}\beta^A_1$ and the
difference will be the insertion of a local functional of 
order $\hbar$. As shown 
in lemma 1 of the appendix,
by adapting a reasoning of \cite{KlZu,VLT} to the present
context of algebraic renormalization, 
the insertion $[\bar s \Xi_1]\circ\G^\infty$ can be replaced by  
$(\G^\infty,{\Xi_1}\circ\G^\infty)+\Delta_c \Xi_1\circ\G^\infty$, and
the difference will again be the insertion of a local functional of 
order $\hbar$.

We thus get 
\be
(\G^\infty,[S_\eta-\hbar\Xi_1\circ\G^\infty])+\Delta_c 
[S_\eta-\hbar\Xi_1\circ\G^\infty]
\nonumber\\
=\hbar\frac{\partial^R \G^\infty}{\partial \xi^A}\beta^A_1
+\hbar^2{\cal B}^\prime\circ\G^\infty.\label{5}
 \ee

Acting with  $(\G^\infty,\cdot)+\Delta_c$ on
(\ref{5}), using (\ref{3}) and $s_Q
{\partial^R\cdot }/{\partial \xi^A}\beta^A_1=0$, we get the consistency
condition $(\G^\infty,{\cal B}^\prime\circ\G^\infty)
+\Delta_c{\cal B}^\prime\circ\G^\infty=0$ so that the reasoning can be
pushed to all orders:
\be
(\G^\infty,[S_\eta-\Xi\circ\G^\infty])+\D_c[S_\eta-\Xi\circ\G^\infty]
=\frac{\partial^R \G^\infty}{\partial \xi^A}\beta^A,\label{68}
\ee
with $\Xi=\Sigma_{n=1}\hbar^n\Xi_n$ and
$\beta^A=\Sigma_{n=1}\hbar^n\beta^A_n$. 

\subsubsection*{Digression}
Let us consider for a moment the following particular 
case. 

(i) All the antibracket maps encoded in $f^A$ are zero so
that $\Delta_c=0$, $\bar s =s=(S(\xi),\cdot)$. 
This happens for instance if the Kluberg-Stern and
Zuber conjecture \cite{KlZu} holds and the BRST cohomology can be
described independently of the antifields.

(ii) The only possibility (for instance for power counting reasons)
for $\Xi_n$ is $\Xi_n=-\gamma_n\int
d^nx\ \phi^*_a\phi^a$, so that $\Xi_n$ is linear in the quantum fields 
and $[s \Xi_n]\circ\ \G^\infty$ can be replaced, according to the
quantum action principle, at each stage in $\hbar$ 
by $(\G^\infty,\Xi_n)$ up to the insertion of a local polynomial 
of higher order in $\hbar$.
Equation (\ref{68}) then takes the more familiar 
form of the Callan-Symanzik
equations in the massless case with anomalous dimension $\gamma
=\Sigma_{k=1}\hbar^k\gamma_k$
for the fields and the antifields \cite{ItZu}:
\be
(\G^\infty,S^\infty_\eta)=
{\partial\G^\infty\over\partial \xi^A}\beta^A,
\ee
with $S^\infty_\eta=\int d^nx\ \phi^*_a(d^{(a)}+\gamma
+x\cdot\partial)\phi^a$,
or explicitly 
\be
\int d^nx\Big[\ {\delta^R \G^\infty\over\delta \phi^a(x)}(d^{(a)}
+\gamma+
x\cdot\partial)\phi^a(x)\nonumber\\+{\delta^R 
\G^\infty\over\delta \phi^*_a(x)}
(n-d^{(a)}-\gamma+x\cdot\partial)\phi^*_a(x)\Big]
={\partial^R\G^\infty\over\partial \xi^A}\beta^A .\label{67}
\ee

\subsubsection*{Remarks on explicit $x$ dependence.}

Note that $S_\eta$ is the generator of the dilatation symmetry 
of the theory. If it corresponds to a non trivial element
of $H^{-1,n}(s|d)$, the question arises whether it should be coupled with a
constant ghost in the extended solution $S(\xi)$ as in \cite{To,Wh}. 
This depends on
whether or not we allow for explicit $x$ dependence in the local
functionals and the cohomology
classes of $s$ we are initially computing and then coupling 
to the solution of the master equation. 

In the previous section, we have supposed that
there is no explicit $x$ dependence in these functionals and  
cohomology classes, because if we 
assume the absence of dimensionful couplings, we cannot control
translation invariance through a corresponding cohomology class, its
generator $S_\mu=\int d^nx\ \phi^*\partial_\mu\phi$ being of dimension
$1$. 

We will assume here that one can apply the 
quantum action principles in the case of an explicit $x$ dependence of
the variation as in (\ref{zin}), at the price of allowing a priori for an 
explicit $x$ dependence of the inserted local functional ${\cal B}$. 
This assumption needs to be
checked by a more careful analysis of the renormalization properties
of the model which is beyond the scope of this paper.

In order to prove 
then that ${\cal B}$ in eq (\ref{zin}) does 
not depend explicitly on $x$, we use 
translation invariance: classical translation invariance 
is expressed through $(S,S_\mu) =0$ with quantum version $(\G^\infty,S_\mu)
=\hbar{\cal D}_\mu\circ\G^\infty$, where the dimension of ${\cal D}_\mu$
is $1$, because there are no dimensionful parameters in the theory. 
Applying $(\cdot,S_\mu)$ to (\ref{zin}), using the graded Jacobi identity 
for the antibracket, the commutation relation $(S_\eta,S_\mu)=-S_\mu$ 
and the result on the dimension of ${\cal D}_\mu$, i.e., the relation
$({\cal D}_\mu,S_\eta)={\cal D}_\mu$, one finds to lowest order 
$({\cal B},S_\mu)=0$. This means that $(\partial_\mu-\partial/\partial x^\mu)
{\cal B}=0$, and since $\partial_\mu {\cal B}=0$, it which shows 
that ${\cal B}$ does not depend explicitly on $x$.

In the general case where we allow for dimensionful couplings
considered below, we will assume that the theory is translation
invariant and that the generator $S_\mu$ is coupled through the
constant translation ghosts $\xi^\mu$. One can then show that the
local cohomology of the BRST operator in form degre $n$ 
for the extended theory can be choosen
to be independent of both $x^\mu$ and $\xi^\mu$ \cite{Whi2}.
In the same way, Lorentz invariance can then be controled 
inside the formalism 
by coupling the appropriate generator.

\subsubsection{Local form of Callan-Symanzik equation}

In order to get non renormalization results for couplings associated to
terms which are invariant only up to boundaries, we need a local
version of the Callan-Symanzik equation. This is easily obtained from
(\ref{4}) by integrations by parts, giving
\be
(S,{L}_\eta)=\partial_\mu J^\mu_\eta,\label{27}
\ee
where we will refer to $J^\mu_\eta$ as the Callan-Symanzik current in
the following. The quantum action principle gives 
\be
(\G^\infty,{L}_\eta)=\partial_\mu
[{J^\mu_\eta}\circ\G^\infty]+ \hbar b_\eta\circ\G^\infty\label{6},
\ee
where $b_\eta$ is a non integrated local polynomial.
We need to know how ${J^\mu_\eta}\circ\G^\infty$ behaves under the
quantum BRST transformations. The classical descent equations obtained
by applying $\bar s$ to (\ref{27}) imply
the existence of $K^{[\nu\mu]}_\eta$ such that 
\be
\bar s J^\mu_\eta=\partial_\nu K^{[\nu\mu]}_\eta.
\ee
Using lemma 1 of the appendix, we get 
\be
(\G^\infty,{J^\mu_\eta}\circ\G^\infty)+\D_c{J^\mu_\eta}\circ\G^\infty
=\partial_\nu
[{K^{[\nu\mu]}}_\eta\circ\G^\infty]+ \hbar b^\mu_\eta\circ\G^\infty,
\ee
where $b^\mu_\eta$ is a non integrated local polynomial.
The consistency condition for (\ref{6}) is then
\be
(\G^\infty,b_\eta\circ\G^\infty)+\D_c b_\eta\circ\G^\infty
+\partial_\mu[b^\mu_\eta\circ\G^\infty]=0,
\ee
giving to lowest order $\bar s b_{\eta}
+\partial_\mu b^\mu_{\eta }=0$.
We want to derive conditions under which, by appropriately modifying
$L_\eta$, $J^\mu_\eta$ and $K^{[\nu\mu]}_\eta$,
this equation reduces to 
\be
(\G^\infty,b_\eta\circ\G^\infty)+\D_cb_\eta\circ\G^\infty=0,\label{25}
\ee
so that to lowest order the cocycle condition is rather 
$\bar s b_{\eta }=0$.

(i) A first possibility would be the absence of non trivial BRST anomalies
in the renormalization of the Callan-Symanzik current, i.e., 
the existence of local, finite, BRST breaking
counterterms $\Sigma^{\mu}_k$ and $\Sigma^{[\nu\mu]}_k$ such that, if
$J^{\infty \mu}_\eta={J^\mu_\eta}-\Sigma_{k=1}\hbar^k \Sigma^{\mu }_k$ and
$K^{\infty[\nu\mu]}_\eta={K^{[\nu\mu]}_\eta}-\Sigma_{k=1}\hbar^k
\Sigma^{[\nu\mu] }_k$, we have
\be
(\G^\infty,J^{\infty \mu}_\eta\circ\G^\infty)+\D_c 
J^{\infty \mu}_\eta\circ\G^\infty=\partial_\nu
[K^{\infty[\nu\mu]}_\eta\circ\G^\infty].\label{10}
\ee
A sufficient condition for this to occur is
$H^{1{+}k,n-1-k}(\bar s|d)=0$, for $k=0,\dots,n-1$. This follows
from an analysis of the quantum descent equations as in
\cite{PiSo}. 
Indeed, the classical descent equations are
\be
\bar s J^{n-1-k}_\eta+dJ^{n-1-k-1}_\eta=0, 
\ee
for $k=0,\dots n-1$
with $J^{-1}_\eta=0$, where we can assume $J^{n-1-k}_\eta\in
H^{k,n-1-k}(\bar s|d)$. The quantum version of the descent equations
are 
\be
(\G^\infty,{J^{n-1-k}_\eta}\circ\G^\infty)
+\D_c{J^{n-1-k}_\eta}\circ\G^\infty
+d[{J^{n-1-k-1}_\eta}\circ\G^\infty]\nonumber\\
=\hbar b_\eta^{n-1-k}\circ\G^\infty. 
\ee
Applying $(\G^\infty,\cdot)+\D_c$, we get the consistency condition 
$(\G^\infty,b_\eta^{n-1-k}\circ\G^\infty)+\D_c
b_\eta^{n-1-k}\circ\G^\infty+
d[b_\eta^{n-1-k-1}\circ\G^\infty]=0$, giving to lowest order 
$\bar s b^{n-1-k}_{\eta }+db^{n-1-k-1}_{\eta }=0$, 
and thus, because of the assumption on the vanishing of the
relevant cohomology classes, $b^{n-1-k}_{\eta }=
\bar s\sigma^{n-1-k}_{\eta 1}+d\sigma^{n-1-k-1}_{\eta
1}$, so that 
\be
(\G^\infty,[{J^{n-1-k}_\eta}-\hbar\sigma^{n-1-k}_{\eta 1}]
\circ\G^\infty)+\D_c([{J^{n-1-k}_\eta}-\hbar\sigma^{n-1-k}_{\eta1})]
\circ\G^\infty)
\nonumber\\+d([{J^{n-1-k-1}_\eta}-\hbar\sigma^{n-1-k-1}_{\eta 1
}]\circ\G^\infty)\nonumber\\
=\hbar^2b_\eta^{\prime n-1-k}\circ\G^\infty,
\ee
and the reasoning can be continued recursively to higher orders, so
that we can achieve (\ref{10}). If we now replace  ${J^\mu_\eta}$ by 
$J^{\infty \mu}_\eta$ in (\ref{6}) at the
expense of modifying $b_\eta$ appropriately, we get the desired
consistency condition (\ref{25}). 

However, the vanishing of the
cohomology groups assumed here is too restrictive to be of
use. Indeed, our aim is to find conditions for the vanishing of the
$\beta$ functions associated to BRST cohomology goups with a non trivial
descent. As shown in the appendix, the local BRST cohomology groups
admit the decomposition $H^{0,n}(\bar s|d)\simeq
lH^{1,n-1}(\bar s|d)\oplus rH^{0,n}(\bar s)$, where the first
group is isomorphic to the local BRST cohomology groups with a non
trivial descent, $H^{0,n}_{nd}(\bar s|d)$. But
$lH^{1,n-1}(\bar s|d)\subset H^{1,n-1}(\bar s|d)$. But by
assumption, this last group vanishes, so that the assumptions
unfortunately imply
the absence of the terms whose non renormalization properties we wanted
to show.

(ii) A second, non trivial case is when 
the Callan-Symanzik current is ``covariantizable'', i.e., when there
exists $J^\mu_{\eta
c}=J^\mu_\eta+\bar s K^\mu_\eta+\partial_\nu
M^{[\nu\mu]}_\eta$ such that 
$\bar s J^\mu_{\eta c}=0$\footnote{As shown in \cite{BBH},
this question can again be formulated as a question of local BRST
cohomology.}. The quantum
version of this equation is 
$(\G^\infty,{J^\mu_{\eta c}}\circ
\G^\infty)+\D_c{J^\mu_{\eta c}}\circ
\G^\infty=\hbar b^\mu_\eta\circ
\G^\infty$
and a sufficient condition for the absence of anomalies for the
covariant Callan-Symanzik current, i.e., the existence of finite 
counterterms to
${J^\mu_{\eta c}}\rightarrow J^{\mu\infty}_{\eta
c}$ such that 
\be
(\G^\infty,{J^{\mu\infty}_{\eta
c}}\circ \G^\infty)+\D_c{J^{\mu\infty}_{\eta
c}}\circ \G^\infty=0,
\ee
is now the vanishing of the strict cohomology 
groups $H^{1,n-1}(\bar s)=0$ without modulo $d$ terms, which is of
course a much less restrictive assumption.

Replacing $L_\eta$ by $L_\eta-\partial_\mu K^\mu_\eta$, we get 
$\bar s (L_\eta-\partial_\mu K^\mu_\eta)=\partial_\mu
J^\mu_{\eta c}$ and the quantum version 
\be
(\G^\infty,L_\eta
-(\partial_\mu K^\mu_\eta)\circ\G^\infty)+
\D_c [L_\eta
-(\partial_\mu K^\mu_\eta)\circ\G^\infty] \nonumber\\
=
\partial_\mu[{J^{\mu\infty}_{\eta
c}}]\circ\G^\infty+\hbar b_\eta\circ
\G^\infty,\label{31}
\ee
with the desired consistency condition (\ref{25}).

{}From $\bar s b_\eta=0$, it follows that $b_\eta=\tilde \beta^jc_j
+\bar s \xi_1$ where $c_j$ is a basis for $H^{0,n}(\bar s)$.
There is however a subtlety here, since there
might be elements of $H^{0,n}(\bar s)$, which become trivial if one
allows for total divergences and could be absorbed by modifying at the
same time $L_\eta$ and the current. In order to deal with this aspect,
we need a decomposition of strict BRST cohomology cocycles
analogous to that of modulo $d$ ones given in the appendix: 
$H^{g,k}(\bar s)=rH^{g,k}(\bar s)\oplus T^{g,k}$, where the space 
$T^{g,k}$ is defined by this equation. In other words,
we split the solutions of the cocycle condition  $\bar s
k=0$ into $k=b^{\bar A}k_{\bar A}
+\bar s l+dm$, where $k_{\bar A}$
are a basis of $rH^{g,k}(\bar s)$, and $\bar s 
m+dn=0$. In this equation, we can assume that $m\in
lH^{g,k-1}(\bar s|d)$, since a trivial solution to this equation can
be absorbed by a redefinition of $l$, and by assumption, 
$m$ can be lifted to give a part of the BRST cocycle $k$. 

Thus, we take 
$b_{\eta}=\beta^{\bar j}_{1\ {\rm
loc} }c_{\bar j
  }+\bar s\xi_1+\partial_\mu m^\mu_1$, so that the
redefinition ${L}_\eta-\partial_\mu K^\mu_\eta\rightarrow 
L^1_\eta={L}_\eta-\partial_\mu K^\mu_\eta-\hbar \xi_1$ 
allows to rewrite (\ref{31}) as
\be
(\G^\infty,L^1_\eta\circ\G^\infty)+\D_cL^1_\eta\circ\G^\infty=\partial_\mu
[J^{\infty \mu}_{\eta c}+\hbar
{m}^\mu_1]\circ\G^\infty\nonumber\\
+\hbar\beta^{\bar j}_{1\ {\rm
loc} }c_{\bar  j}\circ\G^\infty+
\hbar^2b^\prime_\eta\circ\G^\infty.\label{12}
\ee

In order to continue to higher orders, we need to know how the
insertions $m^\mu_1\circ\G^\infty,\beta^{\bar j}_{1\ {\rm
loc} }c_{\bar  j}\circ\G^\infty$ 
behave under quantum BRST transformations. Because
$m^\mu_1\in lH^{0,n-1}(\bar s|d)\simeq H^{-1,n-1}(\bar s|d)$ which
represents the non trivial global currents of the theory
\cite{BBH1}, we assume that those that can appear on the right hand
are covariantizable, as we did for the
dilatation current. Hence, we can replace $m^\mu_1$ by $m^\mu_{1 c}$ at the
expense of modifying $L^1_\eta$ by terms of order at least
$\hbar$. The further assumption $H^{1,n-1}(\bar s)=0$ then
guarantees the existence of counterterms such that
$(\G^\infty,m^{\infty \mu}_{1 c}\circ \G^\infty)+\D_c
m^{\infty \mu}_c\circ \G^\infty =0$. 
{}From lemma 2 of the appendix, it follows 
that there exist finite counterterms $c_{\bar  j}\rightarrow
c^\infty_{\bar  j}$ such that 
\be
(\G^\infty,c^\infty_{\bar  j}\circ\G^\infty)+\D_cc^\infty_{\bar  j}
\circ\G^\infty=\hbar a^i_{\bar j} k^\infty_i\circ \G^\infty, 
\ee
where $k_i\in H^{1,n}(\bar s)$.

We now replace in (\ref{12}) $m^\mu_1$ by $m^{\infty\mu}_{1 c}$ and 
$c_{\bar  j}$ by $c^\infty_{\bar  j}$ and modify $L^1_\eta$ as well as
$b^\prime_\eta$ accordingly. The consistency condition then gives 
to lowest order $\beta^{\bar j}_{1\ {\rm
loc} }a^i_{1 \bar j}k_i=\bar s b^\prime_\eta$ so that 
$\beta^{\bar j}_{1\ {\rm
loc} }a^i_{1 \bar j}=0=\bar s b^\prime_\eta$. 

Going on recursively to higher orders, we find finally
\be
(\G^\infty,L^\infty_\eta\circ\G^\infty)+\Delta_c
L^\infty_\eta\circ\G^\infty
=\partial_\mu
[J^{\infty \mu}_{\eta c}-m^{\infty \mu}_c]\circ\G^\infty
+\beta^{\bar j}_{\rm loc}c^\infty_{\bar  j}\circ\G^\infty,
\ee
where $m^{\infty \mu}_c=\Sigma_{n=1}\hbar^n m^{\infty \mu}_{n c}$
and $\beta^{\bar j}_{\rm loc}=\Sigma_{n=1}\hbar^n \beta^{\bar j}_{n\ 
{\rm loc}}$ and the classical approximation of 
$L^\infty_\eta\circ \G^\infty$ is $L_\eta-\partial_\mu
K^\mu_\eta$. 
We can now integrate to get 
\be
(\G^\infty,S^\infty_\eta\circ\G^\infty)+\D_c S^\infty_\eta\circ\G^\infty=
\hbar\beta^{\bar j}_{\rm loc}\int d^nx\ c^\infty_{\bar  j}\circ\G^\infty,
\ee
where the classical approximation for $S^\infty_\eta \circ\G^\infty$ is the
generator $S_\eta$ of dilatation invariance.
Comparing with (\ref{68}), we thus see that the $\beta$ functions of
the elements of $H_{nd}^{g,n}(\bar s|d)$ vanish and only those
associated to $rH^{g,n}(\bar s)$ may be non vanishing.

\begin{theorem}
Suppose a theory contains only dimensionless coupling constants. 
If (i) the extended BRST symmetry $\bar s$ is
non anomalous, (ii) the Callan-Symanzik current and 
all the non trivial global conserved currents that
can mix with it under renormalization are
covariantizable, and (iii) the renormalization does not introduce BRST
anomalies for these currents, then the 
$\beta$ functions of cohomology classes of $\bar s$ with a 
non trivial descent vanish to all orders.
A sufficient condition for (i) to hold is $H^{1,n}(\bar s|d)=0$, while a
sufficient condition for (iii) to hold is  $H^{1,n-1}(\bar s)=0$.
\end{theorem}

\subsubsection{Application: Pure semi-simple Chern-Simons theory}

Consider semi-simple Chern-Simons theory without matter couplings, 
in a gauge without a dimensionful parameter as for instance the Landau
gauge. 
The minimal solution to the master equation is 
\be
S=\int d^3x\ \epsilon^{\mu\nu\sigma}g_{ij}[
A^i_\mu\partial_{\nu}^jA_{\sigma}+\frac{1}{3} f^j_{kl}A^i_\mu A^k_\nu 
A^l_\sigma] + A^{*\mu}_iD_\mu C^i+\frac{1}{2}C^*_if^i_{jk}C^jC^k.
\label{cs}
\ee
The local BRST cohomology has been worked out for instance
in \cite{BBH2}. 
In ghost number $0$, it
contains only the Chern-Simons terms 
associated to the different simple factors, and all of them involve a non
trivial descent. The remaining local BRST cohomology classes are at least
of ghost number $3$ and do not involve the antifields in a non trivial
way. This means that $\D_c$
vanishes and there is no need to couple different local BRST cohomology
classes to the solution of the master equation to ensure
stability. Thus, we can take the usual BRST differential associated
to the standard solution of the master equation $s=(S,\cdot)$.
Condition (i) is satisfied because 
$H^{1,n}(s|d)=0$. 
It is straightforward to verify that 
$J^\mu_c=0$, the Callan-Symanzik current is trivial. This follows from 
\be
S_\eta=\int d^3x\ A^{*\mu}_i(1+x\cdot\partial)A^i_\mu+
C^*_ix\cdot\partial C^i=(S,M_\eta)+\partial_\mu N^\mu_\eta,
\ee
with $M_\eta=1/4 A^{*\mu}_i A^{*\sigma}_mg^{im}x^\nu\epsilon_{\nu\mu\sigma}
+C^*_ix^\nu A_\mu^i$ and $N^\mu_\eta=A^{*\mu}_ix^\nu A_\mu^i$.

It is a consequence
of the fact that the local BRST cohomology in negative ghost number is
empty, and this both in the space of $x$ dependent and $x$
independent local functionals. There thus are no non nontrivial conserved
currents, so that both assumptions (ii) and (iii) are satisfied.  
Hence, the vanishing of the $\beta$ functions in pure
semi-simple Chern-Simons theory can be traced back completely to 
the structure of the local BRST cohomology of the theory.

\subsection{General broken case}

\subsubsection{Integrated form of Callan-Symanzik equation}

Again, we assume that there are non anomalies, eq. (\ref{3}).
We will now allow for coupling constants $\xi^A$ of 
all possible dimensions $d^{(A)}$ in the theory, 
which could be negative in the case of effective
field theories.
We have 
\be
(L,S_\eta)+{\partial^R L\over\partial \xi^A}
d^{(A)}\xi^A=\partial_\mu (x^\mu L).\label{pc}
\ee
Integrating, one gets 
\be
{\cal C}S=0,\label{77}
\ee
with ${\cal C}=(\cdot,S_\eta)+{\partial^R \cdot\over\partial \xi^A}
d^{(A)}\xi^A$. 
Using (\ref{1}), we get $s_Q {\partial^R\cdot /\partial \xi^A}
d^{(A)}\xi^A=[\bar s,{\cal C}]=0$.
The quantum version of (\ref{77}) is 
\be
{\cal C}\G^\infty=\hbar{\cal B}\circ\G^\infty.
\ee
and, applying $(\G^\infty,\cdot)+\D_c$, the consistency condition to
lowest order implies $\bar s{\cal B}=0$. We can then get as in the
previous section the general form of the integrated 
Callan-Symanzik equation:
\be
(\G^\infty,[S_\eta-\Xi\circ\G^\infty])+\D_c[S_\eta-\Xi\circ\G^\infty]
=\frac{\partial^R \G^\infty}{\partial
\xi^A}(\beta^A-d^{(A)}\xi^A),\label{qcs}
\ee
with $\Xi=\Sigma_{n=1}\hbar^n\Xi_n$ and
$\beta^A=\Sigma_{n=1}\hbar^n\beta^A_n$. 

In the derivation above, we have neglected the fact that there could
be other dimensionful parameters $\alpha^i$ 
in the theory, besides the essential
ones discussed so far. They could for instance come from
the gauge fixation of the theory. Differentiation of the extended
master equation and of $s_Q^2=0$ with respect to $\alpha^i$ and 
using (\ref{8}), (\ref{9}) implies
\be
\frac{\partial^R S}{\partial \alpha^i}=\bar s \Xi_i+
\frac{\partial^R S}{\partial \xi^A}\kappa^A_i,\label{55}
\ee
with 
\be
\frac{\partial^R\cdot }{\partial \xi^A}\frac{\partial^R f^A}{\partial
\alpha_i}
=s_Q \frac{\partial^R\cdot }{\partial \xi^A}\kappa^A_i.
\ee
If $d^i$ is the dimension of $\alpha^i$, we have to add the term 
${\partial^R S}/{\partial \alpha^i}d^{(i)}\alpha^i$ in (\ref{77}). Using
(\ref{55}) and $\D_c S_\eta=0$, (\ref{77}) becomes
\be
\bar s (S_\eta+\Xi_i d^{(i)}\alpha^i)
+\frac{\partial^R S}{\partial
\xi^A}(d^{(A)}\xi^A+\kappa^A_id^{(i)}\alpha^i)=0.
\ee
Together with (\ref{3}) this equation can again be used to prove 
that 
\be
s_Q {\partial^R\cdot }/{\partial
\xi^A}(d^{(A)}\xi^A+\kappa^A_id^{(i)}\alpha^i)=0.
\ee 
The quantum analysis
then proceeds exactly as before, and we get as quantum equation 
(\ref{qcs}) with $S_\eta$ replaced by $S_\eta+\Xi_i d^{(i)}\alpha^i$
and $d^{(A)}\xi^A$ by $d^{(A)}\xi^A+\kappa^A_id^{(i)}\alpha^i$.

\subsubsection{Local form of Callan-Symanzik equation}

To get a local Callan-Symanzik equation, we integrate the first term
in (\ref{pc}) by parts: 
\be
(S,L_\eta)+{\partial^R L\over\partial \xi^A}
d^{(A)}\xi^A=\partial_\mu J^\mu_\eta.\label{50}
\ee
[Again, dimensionful inessential 
parameters can be incorporated by the substitutions
$(S,L_\eta)\rightarrow \bar s (L_\eta+\xi_i d^{(i)}\alpha^i)$, 
$d^{(A)}\xi^A\rightarrow d^{(A)}\xi^A+ \kappa^A_id^{(i)}\alpha^i$ and
finally $J^\mu_\eta\rightarrow J^\mu_\eta-j^\mu_id^{(i)}\alpha^i$,
where 
$$
\frac{\partial^R L}{\partial \alpha^i}=\bar s \xi_i+
\frac{\partial^R L}{\partial \xi^A}\kappa^A_i+\partial_\mu j^\mu_i.]
$$
Let us now assume that there exists $K^\mu_\eta$, $M^{[\nu\mu]}_\eta$
such that if 
\be
J^\mu_{\eta c}=J^\mu_\eta+\bar s K^\mu_\eta
+\partial_\nu M^{[\nu\mu]}_\eta,
\ee
then 
\be
\bar s J^\mu_{\eta c}=0.
\ee
The assumption on the Callan-Symanzik current $J^\mu_\eta$ implies, by
acting with $\bar s$, 
\be
\bar s {\partial^R L\over\partial \xi^A}
d^{(A)}\xi^A=0.
\ee
This is a strong restriction on the terms of the (extended) 
Lagrangian coupled with
dimensionful couplings: differentiation with respect to $\xi^B$ and
putting the $\xi$'s to zero requires all these terms 
to be strictly BRST invariant.
This means that 
the non renormalization theorem to be derived holds only 
if the terms that are invariant up to boundaries are all of
dimension $n$.

Equation (\ref{50}) becomes
\be
\bar s[L_\eta-\partial_\mu K^\mu_\eta]
+{\partial^R L\over\partial \xi^A}
d^{(A)}\xi^A=\partial_\mu J^\mu_{\eta c}.
\ee
The quantum version of this equation is 
\be
[(\G^\infty,\cdot)+\Delta_c][L_\eta-\partial_\mu {K^\mu_\eta}]
\circ \G^\infty+
{\partial^R L\over\partial \xi^A}
d^{(A)}\xi^A\circ\G^\infty
\nonumber\\=\partial_\mu[J^\mu_{\eta c}\circ\G^\infty]
+\hbar b_\eta\circ\G^\infty.
\label{15}
\ee

If the Callan-Symanzik current and ${\partial^R L\over\partial \xi^A}
d^{(A)}\xi^A$ renormalize without anomalies, i.e., if there exist
counterterms such that 
\be
(\G^\infty,J^{\infty\mu}_{\eta c}\circ\G^\infty)
+\D_cJ^{\infty\mu}_{\eta c}\circ\G^\infty=0
\ee
respectively 
\be
(\G^\infty,[{\partial^R L\over\partial \xi^A}
d^{(A)}\xi^A]^\infty\circ\G^\infty)
+\D_c[{\partial^R L\over\partial \xi^A}
d^{(A)}\xi^A]^\infty\circ\G^\infty=0,
\ee
which is guaranteed if both $H^{1,n-1}(\bar s)$ and $H^{1,n}(\bar
s)$ vanish, we can replace $J^{\mu}_{\eta c}$ by
$J^{\infty\mu}_{\eta c}$ and $[{\partial^R L/\partial \xi^A}
d^{(A)}\xi^A]$ by $[{\partial^R L/\partial \xi^A}
d^{(A)}\xi^A]^\infty$ in (\ref{15}) at the expense of modifying $b_\eta$
appropriately. The consistency condition obtained by acting with 
$(\G^\infty,\cdot)+\D_c$ is then the same as in the previous section,
$(\G^\infty,b_\eta\circ\G^\infty)+\D_c b_\eta\circ\G^\infty=0$ and we
are back to the situation studied before.

The final result is 
\be
(\G^\infty,L^\infty_\eta\circ\G^\infty)+\Delta_c
L^\infty_\eta\circ\G^\infty+[{\partial^R L\over\partial \xi^A}
d^{(A)}\xi^A]^\infty\circ\G^\infty
\nonumber\\
=\partial_\mu
[J^{\infty \mu}_{\eta c}-m^{\infty \mu}_c]\circ\G^\infty
+\beta^{\bar j}_{\rm loc}c^\infty_{\bar  j}\circ\G^\infty,
\ee
where again $m^{\infty \mu}_c=\Sigma_{n=1}\hbar^n m^{\infty \mu}_{n c}$
and $\beta^{\bar j}_{\rm loc}=\Sigma_{n=1}\hbar^n \beta^{\bar j}_{n\ 
{\rm loc}}$ and the classical approximation of 
$L^\infty_\eta\circ \G^\infty$ is $L_\eta-\partial_\mu
K^\mu_\eta$. 
Integration gives to  
\be
(\G^\infty,S^\infty_\eta\circ\G^\infty)+\D_c
S^\infty_\eta\circ\G^\infty
+\int d^nx [{\partial^R L\over\partial \xi^A}
d^{(A)}\xi^A]^\infty\circ\G^\infty
\nonumber\\=
\hbar\beta^{\bar j}_{\rm loc}\int d^nx\ c^\infty_{\bar  j}\circ\G^\infty.
\ee
As before, 
the classical approximation for $S^\infty_\eta \circ\G^\infty$ is the
generator $S_\eta$ of dilatation invariance, while the one for 
$\int d^nx [{\partial^R L\over\partial \xi^A}
d^{(A)}\xi^A]^\infty\circ\G^\infty$ is ${\partial^R S\over\partial \xi^A}
d^{(A)}\xi^A$.
Comparing with (\ref{68}), we  see again that the $\beta$ functions of
the elements of $H_{nd}^{g,n}(\bar s|d)$ vanish and only those
associated to $rH^{g,n}(\bar s)$ may be non vanishing.

\begin{theorem}
If (i) the extended BRST symmetry $\bar s$ is
non anomalous and the insertion ${\partial^R L/\partial \xi^A}
d^{(A)}\xi^A\circ\G^\infty$ renormalizes without anomaly, 
(ii) the Callan-Symanzik current and 
all the non trivial global conserved currents that
can mix with it under renormalization are
covariantizable, and (iii) the renormalization does not introduce BRST
anomalies for these currents, then the 
$\beta$ functions of cohomology classes of $\bar s$ with a 
non trivial descent vanish to all orders.
A sufficient condition for (i) to hold is $H^{1,n}(\bar s|d)=0$, while a
sufficient condition for (iii) to hold is  $H^{1,n-1}(\bar s)=0$.
\end{theorem}

\subsubsection{Application: Semi-simple Chern-Simons theory
coupled to matter}

We take as a starting point Chern-Simons theory based 
on a semi-simple Lie algebra to which are coupled matter fields.
The minimal solution to the master equation is given by 
\be
S_{CSM}=S+\int d^3x\ L_M(y^m,D^T_\mu y^m, \dots, D^T_{\mu_1}\dots
D^T_{\mu_k}y^m)
+C^i(T_i y)^m y^*_m,
\ee
with $S$ as in (\ref{cs}). The matter field Lagrangian is supposed 
to be gauge invariant, but the whole theory does not admit additional 
local symmetries. In particular, it is not restricted
to be power counting
renormalizable. 
The matter field Lagrangian admits the decomposition 
$L_M=L^{kin}+k^A K_A$, where the $K_A$ are strictly gauge invariant
polynomials and the $k^A$ some coupling constants. 

The local BRST cohomology \cite{BBH2} in ghost number $0$ 
is exhausted by the Chern-Simons terms
associated to the various simple factors and the different invariant
matter field polynomials. The latter are strictly invariant, while the
former involve a non trivial descent. Neither depend on the antifields
in a non trivial way. The theory itself is thus
stable. The only antifield dependent cohomology classes are in ghost
number $-1$ and are related to the global symmetries of the
theory. Since we are not interested here in controling them, we
will not couple the corresponding generators. Hence, we can assume
$\D_c=0$ and $\bar s = s$.  
We have furthermore: 

(i) the terms with dimensionful couplings are all
strictly invariant, 

(ii) the theory is anomaly free because
$H^{1,n}(s|d)=0$, 

(iii) there is no anomaly in the renormalization of the
invariant terms since $H^{1,n}(s)=0$,

(iv) all the non trivial conserved currents are 
covariantizable \cite{BBH} and $H^{1,n-1}(s)=0$.

Thus, the only thing left to check is that the Callan-Symanzik current
is covariantizable. Because it is not the current of a symmetry
of the theory, we need to modify the reasoning of \cite{BBH} in order
to do so. Let us decompose $S_{CSM}=S_0+\int d^3x\ k^\alpha K_\alpha$,
where $S_0$ is the solution of the master equation
where all the dimensionful $k^\alpha$ of the $k^A$ have been set to
zero. We have 
$(S_0,L_\eta)=\partial_\mu J^\mu_{ 0\eta}$, with 
\be
L_\eta=A^{*\mu}_i(1+x\cdot\partial)A^i_\mu+
C^*_ix\cdot\partial C^i+y^*_m(d^{(m)}+x\cdot\partial)y^m,
\ee
so that dilatation is a
true global symmetry of the theory. It follows from \cite{BBH} that
both the generator and its current can be covariantized. Explicitly, if
$M= C^*_i x^\nu A_\nu^i$ and $N^\mu=A^{*\mu}_ix^\nu A_\nu^i$, we have 
\be
L^c_\eta=L_\eta-(S_0,M)-\partial_\mu N^\mu \nonumber\\
= A^{*\mu}_i x^\nu F_{\nu\mu}^i
+y^*_m d^{(m)}y^m+ y^*_m x^\nu D^T_\nu y^m,
\ee
with $(S_0,L^{\eta c})=\partial_\mu J^\mu_{0 \eta c}$ and 
$(S_0,J^\mu_{0 \eta c})=0$. 

On the other hand, we have  
$(S_{CSM},L_\eta)+ K_\alpha d^{(\alpha)}k^\alpha= \partial_\mu
J^\mu_\eta
$. 
Furthermore, $L_\eta-(S_{CSM},M)-\partial_\mu N^\mu=L^c_\eta$ because 
$(\int d^3x\ k^\alpha K_\alpha, M)=0$. This means that
$(S_{CSM},L^c_\eta)+K_\alpha d^{(\alpha)}k^\alpha=
\partial_\mu 
[J^\mu_\eta-(S_{CSM},N^\mu)]
$. As explained in
\cite{BBH}, it follows that one can replace
$J^\mu_\eta-(S_{CSM},N^\mu)$ by $J^\mu_{\eta c}$ with
$(S_{CSM},J^\mu_{\eta c})=0$. 

This proves the vanishing of the $\beta$ functions associated to the
Chern-Simons terms for this power counting non renormalizable 
model, independently of the choice of the
gauge fixing fermion. Note that invariant terms
containing covariant derivatives of
the Yang-Mills field strength, which are all cohomologically trivially
can be incorporated by coupling them with inessential parameters as
discussed in detail in \cite{LLSVV}. The covariant version of the
generator of dilatation is also crucial in \cite{CFHP}, where it has
been obtained by using the trace of the improved energy-momentum tensor.

\section*{Conclusion}

To summarize, the main idea of the paper is that the terms associated
to the $\beta$ functions can be considered as anomalies for (broken)
dilatation invariance and as such, they have to satisfy consistency
conditions. It is possible to formulate a local version of the
Callan-Symanzik equation, where, under suitable assumptions, 
these conditions require the
dilatation anomalies to be strictly invariant, and not only invariant
up to a total divergence, as follows from the usual integrated
Callan-Symanzik equation. 

The general conditions for the validity of the non renormalization
theorem given here can of course be checked in a
straightforwrd way for other (topological) models than 
Chern-Simons theory, as soon as the local BRST cohomology of the
theory is known. 

The main assumption for the non renormalization mechanism given here 
is the non anomalous renormalization of the Callan-Symanzik
current. If this current is not covariantizable, i.e., if it does not
admit a representative which has no non trivial descent, the anomalies
in the renormalization of this current are characterized by
$H^{1,n-1}(\bar s|d)$ and this group will not be vanishing
if there are terms of ghost number $0$ and form degree $n$ involving a
non trivial descent.

Trying to understand the non 
renormalization theorems for the axial or the
non abelian anomaly or the ones in supersymmetric theories by similar
cohomological techniques will be the object of future investigations. 

\section*{Acknowledgements} 
The author wants to thank F.~Brandt, J.~Gomis, 
P.A.~Grassi, M.~Henneaux, T.~Hurth, O.~Piguet, S.P.~Sorella, 
A.~Wilch and S.~Wolf
for useful discussions. He acknowledges the hospitality of 
the Erwin Schr\"odinger International Institute for Mathematical 
Physics in Vienna and of the Department for Theoretical Physics of the 
University of Valencia, where this work has been completed.

\section*{Appendix}

\begin{lemma}
In the anomaly free case defined by (\ref{3}), the equation 
\be
[\bar s \Xi]\circ\G^\infty=(\G^\infty,\Xi\circ\G^\infty)
+\Delta_c[\Xi\circ\G^\infty]
+\hbar{\cal D}\circ\G^\infty
\ee
holds, for local functionals ${\cal D}$ and $\Xi$.
\end{lemma}

\proof{Consider $S_\rho={S}^\infty+\Xi\rho$. Then
$\frac{1}{2}(S_\rho,S_\rho)+\Delta_c S_\rho =\bar s\Xi\rho
+O(\hbar)+O(\rho^2)$.
The quantum action principle applied to this equation gives
$\frac{1}{2}(\G_\rho,\G_\rho)+\Delta_c\G_\rho=
{\cal D}(\rho)\circ\G_\rho$.
Putting $\rho$ to zero and using (\ref{3}), we get ${\cal D}(0)=0$, so
that ${\cal D}(\rho)={\cal D}^\prime(\rho)\rho$. If we now differentiate
with respect to $\rho$ and set $\rho$ to zero, we get
$(\G^\infty,\Xi\circ\G^\infty)+\Delta_c[\Xi\circ\G^\infty]
=\Delta^\prime(0)\circ\G^\infty$. To
lowest order in $\hbar$, we find 
$\Delta^\prime(0)=\bar s \Xi$.}

A different proof of this lemma can be obtained by using the so called
extended BRST technique \cite{PiSi,LLSVV}.
\vspace{.5cm}

\noindent{\bf Decomposition
of BRST modulo $d$ cohomology classes:} \cite{DHTV}

A BRST cocycle modulo $d$ is given by a $k$-form $b$ of ghost number $g$
satisfying $\bar s b+dc=0$ for some $c$. The equivalence classes
$[b]$ under 
the equivalence relation $b\sim b+\bar s f + d g$ are the elements of
$H^{g,k}(\bar s|d)$. 
The descent equations, following from the triviality of the cohomology
of $d$ in form degree less than $n$, 
imply that $\bar s c+de=0$ for some $e$,
so that $[c]\in H^{g+1,n-1}(s|d)$. Consider the map ${\cal D}:
H^{g,n}(\bar s|d)\longrightarrow H^{g+1,n-1}(\bar s|d)$ defined by
${\cal D}[b]=[c]$. It is straightforward to verify that (i) this map
is well defined, i.e., it does not depend on the choice of representatives,
(ii) $Ker\ {\cal D}$ is determined by $[b]$ with $\bar s b=0$; this
space will be denoted by $rH^{g,k}(\bar s)$ and corresponds to those
elements of $H^{g,k}(\bar s)$ which remain non trivial under the
more general $\bar s$ modulo $d$ coboundary condition, and (iii) $Im\
{\cal D}$ is given by those elements $[c]\in
H^{g+1,k-1}(\bar s|d)$, which can be lifted; we denote the
corresponding subspace by $lH^{g+1,k-1}(\bar s|d)$. 

We thus have the
direct sum decomposition $H^{g,k}(\bar s|d)=rH^{g,k}(\bar s)\oplus
H_{nd}^{g,k}(\bar s|d)$, where $H_{nd}^{g,k}(\bar s|d)\simeq
lH^{g+1,n-1}(\bar s|d)$. The subspace $H_{nd}^{g,k}(\bar s|d)$ is
the subspace of cohomology classes with a non trivial descent.

\begin{lemma}
In the anomaly free case defined by (\ref{3}), 
any integrated BRST cohomology class $K_\alpha=\int d^nx\ k_\alpha$ in ghost
number $g_\alpha$, $k_\alpha\in H^{g_\alpha,n}(\bar s|d)$, 
there exist counterterms 
$K_\alpha\rightarrow
K_\alpha^\infty$ such that 
\be
(\G^\infty,K^\infty_\alpha\circ\G^\infty)+\Delta_c
K^\infty_\alpha\circ\G^\infty
=\hbar a^\beta_{\alpha
  }K^\infty_\beta\circ\G^\infty,
\ee
where $k_\beta\in H^{g_\alpha+1,n}(\bar s|d)$ and
$a^\beta_{\alpha}a^\gamma_{\beta}=0$ as a power series in $\hbar$. 

The same result holds for non integrated strict BRST cohomology classes 
$k_\alpha\in H^{g_\alpha,k}(\bar s)$.
\end{lemma}

\proof{We have $\bar s K_\alpha=0$ so that according to lemma 1, 
$(\G^\infty,K_\alpha\circ\G^\infty)+\D_c K_\alpha\circ\G^\infty=\hbar
{\cal A}_\alpha\circ\G^\infty$.
The consistency implies to lowest order that
${\cal A}_{\alpha }=a^\beta_{\alpha
  1}K_\beta+\bar s \Sigma_{1\ \alpha}$, with $k_\beta\in 
H^{g_\alpha+1,n}(\bar s|d)$. Hence, through the redefinition
$K^1_\alpha=K_\alpha-\hbar\Sigma_\alpha^1$, we can achieve 
\be
(\G^\infty,K^1_\alpha\circ\G^\infty)+\D_cK^1_\alpha\circ\G^\infty
=\hbar a^\beta_{\alpha 1}K^1_\beta\circ\G^\infty
+\hbar^2{\cal A}^\prime_\alpha\circ\G^\infty.
\ee
To lowest order, the consistency condition for this equation gives 
$a^\beta_{\alpha
  1}a^\gamma_{\beta 1}K_\gamma=\bar s {\cal A}^\prime_\alpha$,
implying $a^\beta_{\alpha 1}a^\gamma_{\beta1}=0=\bar s {\cal
A}^\prime_\alpha$. The reasoning can be pushed recursively to
higher orders in $\hbar$, with the announced result. The proof for the
strict cohomology classes proceeds in the same way.}

\vfill
\pagebreak


\end{document}